\documentclass{article}

\usepackage[T1]{fontenc} 
\usepackage{float}       
\usepackage{helvet}      
\usepackage{mathptmx}    
\usepackage{etoolbox}
\usepackage{setspace}    

\usepackage{bm}
\usepackage{graphicx}
\usepackage{amssymb}
\usepackage{amsmath}

\usepackage{caption}
\usepackage{fancyhdr}
\usepackage{float}

\begin{document}

\centerline{\bf Phonon Interactions in Metal Halide Perovskites elucidated by Raman Scattering}

\vspace{0.5cm}

\centerline{Alejandro R.~Go\~ni*$^{1,2}$}

\vspace{0.5cm}

\noindent
$^{1}$Institut de Ci\`encia de Materials de Barcelona, ICMAB-CSIC, Campus UAB, 08193 Bellaterra, Spain

\noindent
$^{2}$ICREA, Passeig Llu\'is Companys 23, 08010 Barcelona, Spain\\
*Email: goni@icmab.es

\vspace{1cm}

\begin{abstract}
There is a growing consensus that the exceptional optoelectronic properties of metal halide perovskites (MHPs) are largely due to the peculiar interplay between the inorganic cage lattice, composed of a labile network of corner-sharing metal halide octahedra, and the A-site cationic sublattice. This interaction significantly affects the vibrational spectrum of MHPs (phonon frequencies, linewidths, and lifetimes), resulting from the effects of lattice potential anharmonicity and/or static/dynamic disorder. Raman scattering is a suitable technique to probe phonon interactions in solids, allowing for the in-situ characterization of chemical environments, revealing the nature of lattice vibrations. In this perspective, the available experimental evidence of the aforementioned interplay will be reviewed with special emphasis on understanding Raman signatures depending on whether the coupling is principally mediated by hydrogen bonding or steric hindrance. The controversy about the origin of a strong Raman background, steeply rising towards zero Raman shift and called \textit{central peak}, will be specifically addressed. This background signal, which is typically observed in the temperature range of stability of cubic and tetragonal phases when the A-site cation dynamics unfold, will be shown to be mostly due to disorder-induced second-order acoustic-phonon Raman scattering. This interpretation receives support from other semiconductor systems with nanoscale structural disorder, where the central Raman peak arises either from the vertical misalignment of Ge quantum dots in multi-stack heterostructures or from the interface roughness exhibited by short-period GaAs/AlAs superlattices. In this way, a unifying picture of phonon interactions in MHPs and how they impact different Raman processes is provided, which is key to interpreting their Raman spectra.           
\end{abstract}

\newpage

\section{\label{Intro}Introduction}

The physical properties of solids (electronic, optical, thermal, transport, vibrational, etc.) are primarily determined by the crystal structure, which, in turn, results from the nature of the atomic bonding (covalent, ionic, metallic, or steric) and the need for charge neutrality and stoichiometric balance. In this respect, metal halide perovskites (MHPs) with formula ABX$_3$, being B a divalent metal (Pb$^{2+}$ or Sn$^{2+}$), X a halide anion (I$^{-}$, Br$^{-}$, Cl$^{-}$) and A either organic molecules like methylammonium (MA$^{+}$) and formamidinium (FA$^{+}$) or an inorganic cation (Cs$^{+}$), are pretty unique. Their structure consists in a perovskite-like inorganic but labile network of corner-sharing metal-halide BX$_6^-$ octahedra with the A-site cations filling the inorganic-cage voids, forming the second sublattice that compensates the net charge to zero. A prominent feature of MHPs is the peculiar interplay between the organic and inorganic degrees of freedom, which plays a crucial role in their structural, optical, as well as transport properties and, most importantly for this work, in their vibrations \cite{ivano16a,guoxx17a}. Interestingly, the A-site cations are free to move (translate, rotate and librate) inside the cage voids, exhibiting an uncoordinated roto-translational dynamics. This dynamics is fully or partially (in-plane) unfolded in the cubic and tetragonal phases, respectively, whereas in less symmetric orthorhombic phases the A-site cations are locked in certain positions and orientations inside the voids \cite{frost16a}. This is an experimentally and theoretically well-established fact either directly assessed by ultra-fast vibrational spectroscopy \cite{bakul15a,selig17a} or indirectly inferred from the analysis of the atomic displacement parameter in neutron scattering \cite{welle15a,weber18a} and X-ray diffraction experiments \cite{szafr16a}. The A-site cation dynamics is also theoretically well accounted for within molecular-dynamics calculations \cite{ghosh17a,ghosh19a,maity23a}. For an insightful perspective on the effects of the A-site cation dynamics on the material properties of MHPs, and particularly on phonons, see the work of Gallop et al. \cite{gallo18a}. However, subsequent experimental and theoretical results have shed new light on several issues related to phonon interactions, which will be addressed in this work. Furthermore, here only phonon interactions in three-dimensional (3D) MHPs will be considered; for a recent review on Raman spectroscopy in 2D layered hybrid perovskites see the work of Spirito et al. \cite{spiri22a}.  

Raman spectroscopy is a fast, contactless, and non-invasive technique whose power lies primarily in its sensitivity for the in-situ characterization of chemical environments in materials and for revealing the nature of crystal lattice vibrations (phonons) \cite{yuxxx95a}. This is so mainly because the characteristic frequencies of the phonon modes observed in the Raman spectrum is a vibrational fingerprint of the material. However, other features such as Raman linewidths are used much less frequently for materials characterization, but they also contain valuable information about structural inhomogeneities and lattice anharmonicity. In the case of MHPs, it will be shown that a close inspection of the changes in the Raman linewidths of the inorganic cage phonons as a function of temperature, pressure or halide and/or A-site cation composition allows for drawing important conclusions about the nature of the mentioned interplay between inorganic cage lattice and A-site cation network \cite{gonix24a}. For example, around room temperature or higher, when the A-site cation dynamics is fully unleashed, the concomitant dynamic disorder is at the origin of a strong inhomogeneous broadening observed in the Raman spectra of the inorganic cage phonons. Conversely, locking the A-site cations in fixed positions and/or orientations inside the cage voids either by lowering the temperature or increasing an externally applied pressure leads to a marked sharpening of the Raman peaks in the absence of dynamic disorder. In this case, however, a coupling (H-bonding or Coulomb type) remains between lattice vibrations and A-site cations, giving rise to large lattice anharmonicities that impact the homogeneous Raman linewidths. Another somehow controversial issue concerns the origin of a strong Raman background, steeply rising towards zero Raman shift, thus called \textit{central peak} \cite{yaffe17a}, which is typically observed in the temperature range of stability of cubic and tetragonal phases when the A-site cation dynamics unfolds. This Raman signature will be shown to be mostly due to disorder-induced second-order acoustic-phonon Raman scattering; interpretation supported by the observation of the same phenomenology in other semiconductor systems with nanoscale structural disorder, where the central Raman peak arises either from the vertical misalignment of Ge quantum dots in multi-stack heterostructures \cite{lacha07a,alvar08a} or from the interface roughness exhibited by short-period GaAs/AlAs superlattices \cite{rufxx94a,belit94a}.       

\section{\label{R&D}Results and Discussion}
\subsection{\label{Interplay}Interplay between Inorganic Cage and A-Site Cations}

\begin{figure}[H]
  \includegraphics[width=11cm]{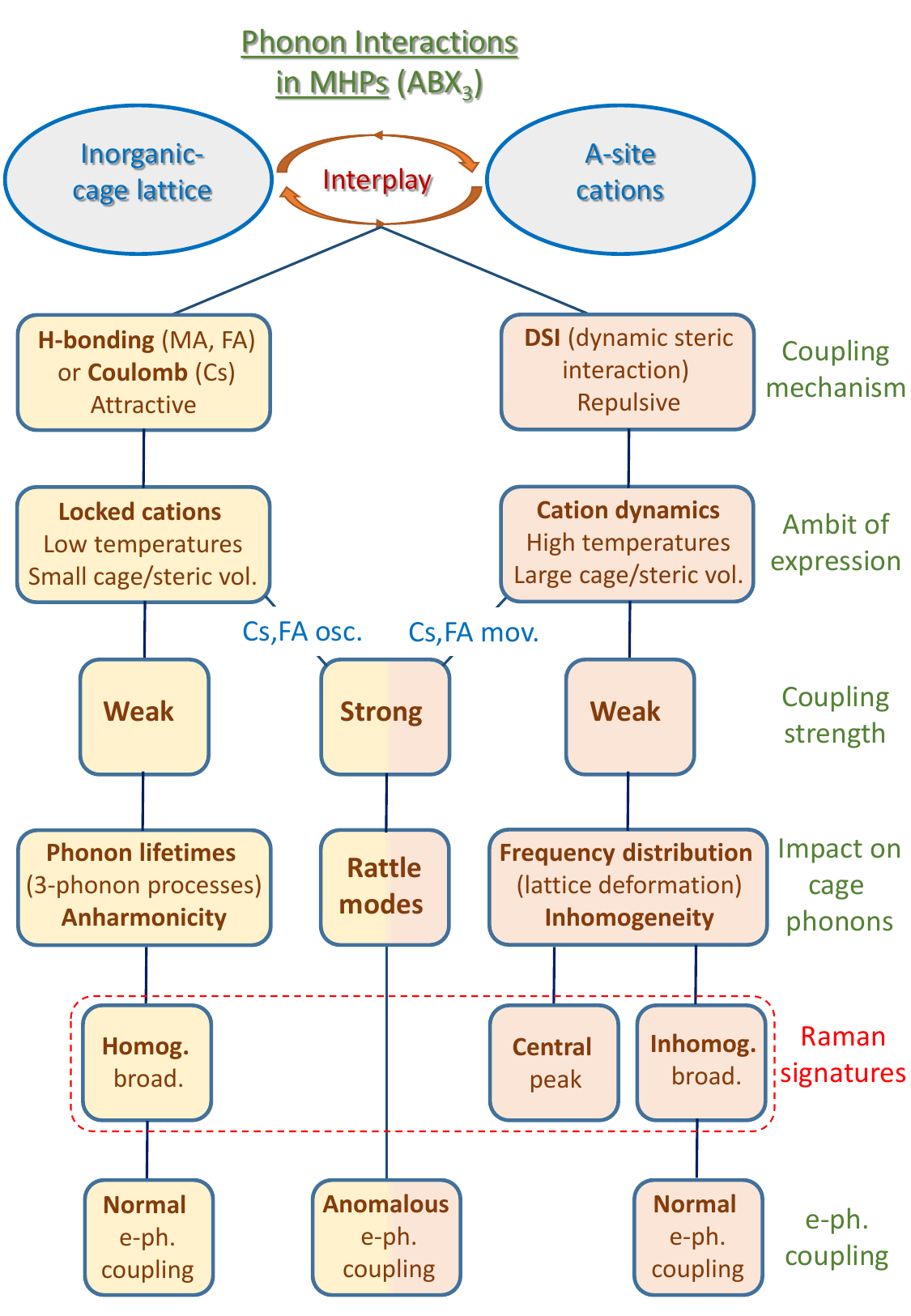}
  \caption{Sketch summarizing how the interplay between inorganic lattice and A-site cations determines phonon interactions in MHPs with formula ABX$_3$, leading to different Raman scattering signatures depending upon the nature of the coupling.
  \label{PhonInter}}
\end{figure}

Before addressing the main topic of this work, namely the elucidation of different types of phonon interactions in MHPs using Raman scattering, it is very instructive to analyze the physics underlying the influence that the interaction between inorganic cage phonons and the A-site cation dynamics exerts on the former, depending on the nature of the coupling. The diagram shown in Fig. \ref{PhonInter} will serve as a guide for this analysis. The interplay between the inorganic cage lattice and the network of A-site cations picks up contributions from two interactions with different origin and acting at different length scales: Hydrogen bonding and steric effects. Hydrogen bonding results from the electrostatic interaction between the hydrogen atoms of the organic cations (mainly of the amine group) and the negatively charged halide anions. In the case of the Cs$^+$ cations, H bonding is replaced by the bare electrostatic (Coulomb) anion-cation attraction. H bonding is ubiquitous in hybrid halide perovskites and has been repeatedly invoked to explain the structural phase behavior of MA,FA lead halides as a function of temperature \cite{yinxx17a,maity23a}, pressure \cite{capit17a,yesud20a} or halide substitution \cite{konto20a}. In contrast, steric effects corresponds to non-bonding dipole-dipole interactions between molecules and/or atoms, which are well described by a Lennard-Jones potential. At large distances steric effects correspond to the weak van der Waals attraction that is much weaker than electrostatic interactions, being thus negligible against H bonding.  However, at short distances the repulsion between the electronic clouds of neighboring atoms or molecules comes into play and the steric interaction becomes strongly repulsive. In the case of MHPs, steric effects are intimately related to the movement of the A-site cations inside the cage voids, which provides the necessary kinetic energy to bring cations and anions sufficiently close together. Hence, the steric repulsion will be hereafter called dynamic steric interaction (DSI).

\begin{figure}[H]
  \includegraphics[width=12cm]{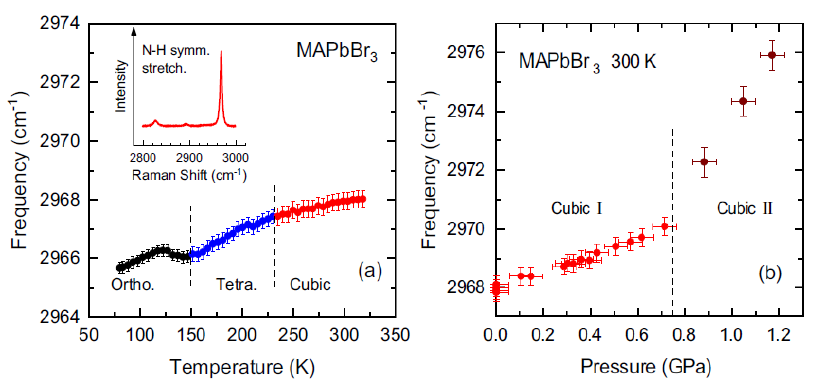}
  \caption{The frequency of the N–H symmetric stretching vibration of the MA cations in MAPbBr$_3$ (a) as a function of temperature at ambient pressure and (b) as a function of pressure at room temperature. The pressures at which the phase transitions occur are marked with vertical dashed lines and the different phases are indicated. The inset shows a representative Raman spectrum in the range of the N–H stretching vibrations around 3000 cm$^{-1}$ (Adapted from K. Xu et al., Sci. Rep., Vol. 13, 9300, 2023; licensed under a Creative Commons Attribution (CC BY) license \cite{xuxxx23a}).   \label{DSI-H-bond} }
\end{figure}

Apart from contributing to the structural stability of the low-temperature orthorhombic phases of MHPs, first-principle calculations have shown that the inorganic-cage/cationic-network interaction is instrumental for the tilting of the PbX$_6$ octahedra.\cite{leexx16a,leexx16b} Furthermore, molecular dynamics (MD) simulations have revealed a one-to-one relationship between the roto-translational dynamics of molecular cations and both octahedral tilting and local structural deformations \cite{ghosh17a,ghosh19a,maity23a}. This is the origin of the dynamic disorder caused by unleashed A-site cation dynamics. Interestingly, a recent high-pressure Raman study on MAPbBr$_3$ has clearly demonstrated that at the temperatures for which the A-site cation dynamics is unfolded, dynamic disorder is principally mediated by DSI rather than H bonding \cite{xuxxx23a}. One comes to this conclusion after close inspection of the temperature and pressure dependence of the N-H symmetric stretching vibration ($\nu_s$) of the MA cations in MAPbBr$_3$, as determined by Raman scattering (strong peak centred at around 3000 cm$^{-1}$, shown in the inset to Fig. \ref{DSI-H-bond}a). This vibrational mode provides direct information about the coupling between the inorganic cage and the A-site cations, specifically regarding the relative importance of H bonding and steric hindrance. Its frequency increases or decreases with changes in temperature, pressure, or composition, depending on whether the coupling between the two sublattices is dominated by steric or hydrogen bonding effects, respectively. The frequency $\nu_s$ is determined by the strength of the covalent N-H bond. In the H-bonding case, the electrostatic attraction between the H$^+$ and the negative halide ion of the inorganic cage weakens the N-H bond by elongating it, thus causing a redshift \cite{wolff90a}. On the contrary, the DSI is repulsive and stronger the closer the H and the halide atom become, which in turn shortens the N-H bond, causing a blueshift. As shown in Fig. \ref{DSI-H-bond}a, $\nu_s$ decreases slightly with decreasing temperature (about 2 cm$^{-1}$ from room temperature down to 80 K), which means that H-bonding increases in importance with decreasing temperature, while the gradual cooling of the MA dynamics diminishes the steric effects. In fact, in the low-temperature orthorhombic phase, H-bonding is crucial to determine the arrangement (position and orientation) of the MA molecules within the cage voids \cite{yixxx22a}. However, at room temperature the application of moderate pressure causes a strong increase in frequency of the N-H stretching mode (ca. 8 cm$^{-1}$ up to 1.2 GPa), as displayed in Fig. \ref{DSI-H-bond}b. This is compelling evidence that, when the MA dynamics is unfolded, the DSI dominates the inter-sublattice coupling. The prominent role of DSI at ambient conditions is also theoretically predicted and experimentally assessed by the blueshift of $\nu_s$ for a reduction of the lattice parameter by the substitution of the halide atom from I to Br to Cl \cite{leguy16a}. Further support for this idea comes from a recent study that combines Raman scattering and density functional theory (DFT), where the entire vibrational spectrum of {\it isolated} MA$^+$ and FA$^+$ molecules is compared with that of MAPbI$_3$ and FAPbBr$_3$, respectively \cite{ibace22a}. This comparison clearly shows that there are no hydrogen bonds in MHPs at room temperature.

In short, the ambits of expression of H-bonding and DSI are quite distinct. H-bonding is the dominant inter-sublattice coupling mechanism when the A-site cation dynamics is frozen and the cations are locked inside the cage voids with totally restricted roto-traslational degrees of freedom. This situation typically occurs in the less-symmetric orthorhombic phases, i.e., at low temperatures and/or when there is little space for movement due to a small cage-void volume ($V_V$) compared to the steric volume ($V_A$) of the A-site cation. On the contrary, the DSI is the most relevant interaction at high temperatures (around ambient and higher) in the stability range of the tetragonal and cubic phases, for which the A-site cation dynamics is partially or totally unfolded, respectively. This is so because the strength of the DSI increases with the amount of dynamic disorder which is determined by the ratio $V_V/V_A$ \cite{xuxxx23a}, rather than the tolerance factor. In fact, these phases that exhibit greater symmetry and larger $V_V/V_A$  ratios, are actually stabilized by cationic motion, which acts against (static) octahedral tilting, thus favoring volumetric cubic and planar tetragonal A-site environments.  

The impact of the interaction between both sublattices on the inorganic cage phonons also depends on the coupling strength. If the coupling is sufficiently strong, a new type of mixed vibrational modes appears in the cage phonon density of states. These modes correspond to combinations of specific cage phonons, mainly involving octahedral tilting, synchronized with some translational, rotational, or librational degree of freedom of the A-site cations. Due to their nature, these modes are called \textit{rattlers} and in CsPbBr$_3$, for example, Cs rattlers are considered to be responsible for the extremely low lattice thermal conductivity of the material, according to MD calculations \cite{lahns24a}. Recently, Cs rattlers have also been invoked to explain the atypical temperature slope of the fundamental gap of low-content, Cs-substituted Cs$_x$MA$_{1-x}$PbI$_3$ single crystals \cite{perez23a} as well as CsPb(Br,Cl)$_3$ nanocrystals with high Cl concentrations \cite{fasah25b}. The rattlers were associated to the dynamic tilting of the PbI$_6$ octahedra in synchrony with either the translational dynamics of the Cs cations between equivalent potential minima of the cage voids or with the Cs oscillations in the potential well of the voids, respectively. In both cases, the Cs rattlers led to the activation of an anomalous electron-phonon interaction mechanism with opposite sign compared to the normal electron-phonon coupling term related to the inorganic cage phonons \cite{franc19a,fasah25a}. In the same way, FA rattler-mediated anomalous electron-phonon coupling is considered to be at the origin of the atypical and pronounced temperature-induced gap bowing observed in FA$_x$MA$_{1-x}$PbI$_3$ mixed-cation single crystals \cite{franc20a} at intermediate FA compositions in the temperature range of stability of a presumably tetragonal phase \cite{xuxxx26a}. Conversely, in the weak coupling limit, the effect of the inorganic-cage/cationic-network interaction on the phonon spectrum of MHPs differs in nature depending on whether the coupling mechanism is H bonding or DSI-mediated, leading to homogeneous or inhomogeneous broadening of the Raman peaks, respectively (see Fig. \ref{PhonInter}), as discussed in detail below. 

\subsection{\label{Linewidths}Raman Linewidths: Homogeneous vs Inhomogeneous Broadening}

Paradoxically, not much information about the interaction between the inorganic cage and the cationic network can be extracted from the Raman frequencies, compared to the linewidths, for example. However, for the sake of completeness, a (non-exhaustive) list will be given of works in which a Raman mode assignment is performed by comparing measured Raman frequencies with the results of first-principles lattice dynamics calculations for various hybrid perovskites, namely: MAPbI$_3$ \cite{brivi15a,ivano16a,perez18a,sharm20a}, MAPbBr$_3$ \cite{yinxx17a}, MAPbX$_3$ (X=I, Br, Cl) \cite{leguy16a}, FAPbX$_3$ (X=I, Br, Cl) \cite{konto20a}, FAPbI$_3$ \& APbBr$_3$ (A=FA, Cs) \cite{ibace22a}. In all cases, the mode assignment has been performed, when applies, for both the inorganic cage phonons with frequencies below 200 cm$^{-1}$ and the molecular vibrations of the organic cations whose frequencies are in the range from 200 cm$^{-1}$ to ca. 3500 cm$^{-1}$; the former employing results from high-resolution, low-temperature Raman measurements.

Generally speaking and assuming that the resolution of the Raman spectrometer is sufficiently high, such that instrumental broadening can be neglected, the linewidths of the Raman peaks are basically determined by two contributions: The {\it homogeneous} broadening, given by the inverse of the phonon lifetime, and the {\it inhomogeneous} broadening, which arises a from any inhomogeneity in the sample leading to a local distribution of bond lengths. At the origin of the homogeneous linewidth is just the uncertainty principle which states that the uncertainty in energy (phonon linewidth) increases in inverse proportion to the uncertainty in time (phonon lifetime). Time-dependent quantum mechanics tells us that the lineshape of homogeneously broadened Raman peaks is a Lorentzian function. A finite lifetime implies that there is a certain perturbation that causes a sizable anharmonicity in the vibrational (phononic) spectrum of the material considered. In good quality crystals, anharmonicities are mainly due to phonon-phonon scattering (three-phonon processes). Since the strength of the phonon-phonon interaction increases with phonon population, the homogeneous linewidth $\Gamma$ is strongly temperature dependent. The peculiarity of MHPs is that the coupling via H bonding, even if the A-site cations are locked inside the cage voids, is primarily affecting the lifetime of the cage phonons, thus increasing lattice anharmonicity. The reduction in phonon lifetime directly translates into an enhanced homogeneous line broadening. The latter is more pronounced the higher the temperature and, more importantly, the smaller the cage voids in comparison to the steric bulk of the A-site cations \cite{xuxxx23a}.

In contrast, inhomogeneous linewidths have a totally different origin. The phonon frequencies are largely determined by the characteristics of the crystal lattice bonds. Any inhomogeneity in crystal structure that locally generates strain or lattice deformation, would also induce inhomogeneous broadening of the phonons. The distribution of bond lengths, i.e. of bond strengths, created by the inhomogeneity leads to a distribution of phonon frequencies $\Delta \omega$. A "normal" distribution of frequencies produces a Gaussian lineshape in the Raman spectrum with a full width at half maximum (FWHM) given by the inhomogeneous broadening, provided that $\Delta \omega>\Gamma$. The Gaussian lineshape corresponds to the convolution of the Lorentzian peaks representing the normal distribution of frequencies. Since such distribution does not change with temperature, inhomogeneous linewidths are temperature insensitive. For MHPs the dynamic disorder induced by the roto-translational A-site cation dynamics leads to a distribution of phonon frequencies due to local deformations and concomitant tilting of the BX$_6$ octahedrons. This, of course, results in strongly inhomogeneously broadened Raman peaks. This is so because the typical time scale of the A-site cation translations and/or jump-like rotations is of several ps, whereas phonon lifetimes are always in the sub-ps range \cite{frost16a}. In fact, during the lifetime of any cage phonon, the A-site cation network is {\it static}. The same occurs with the associated lattice inhomogeneity corresponding to the ensemble of local deformations induced by the different positions and/or orientations adopted by the cations inside the cage voids. A Raman measurement can thus be considered as a sort of stroboscopic sampling of the set of lattice deformations, updated every couple of ps according to the A-site cation dynamics, leading to a distribution of bond lengths causing the inhomogeneous broadening of the cage phonons.

\begin{figure}[H]
  \includegraphics[width=13cm]{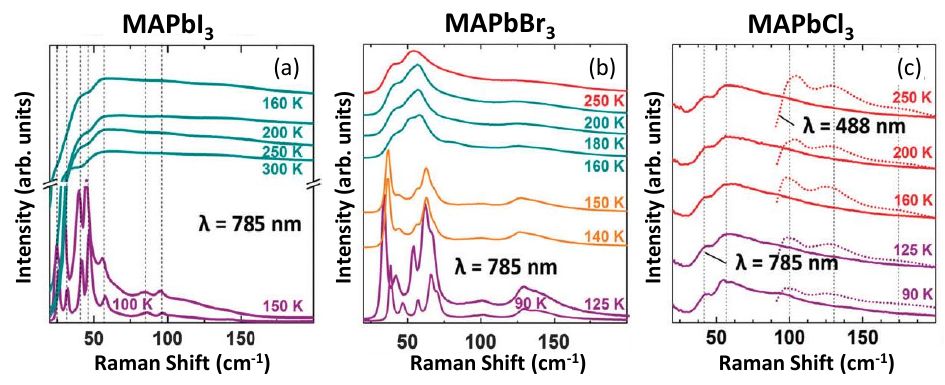}
  \caption{Temperature dependence of the Raman spectra of (a) MAPbI$_3$, (b) MAPbBr$_3$ and (c) MAPbCl$_3$ in the spectral range of the inorganic cage phonons. In (a), the case of MAPbI$_3$, a logarithmic scale is used above the y-axis break. Different colors represent different crystal phases: The orthorhombic phase in purple, the tetragonal phases in orange and dark cyan and the cubic phase in red (Reproduced with permission from Phys. Chem. Chem. Phys. 18, 27051 (2016). Copyright 2016 Royal Society of Chemistry \cite{leguy16a}).
  \label{hom-inhom broad}}
\end{figure}

Figure \ref{hom-inhom broad} displays a series of Raman spectra measured in the spectral region below ca. 200 cm$^{-1}$ of the inorganic cage modes at different temperatures for the three lead halide perovskites MAPbX$_3$ with X=Cl, Br and I \cite{leguy16a}. The different line colors indicate the different crystalline phases adopted by the perovskites at different temperatures, where purple represents the orthorhombic phase, orange and dark cyan the tetragonal ones and red the cubic phase. These spectra exhibit a striking systematic concerning the linewidths of the Raman peaks associated with the inorganic cage phonons as a function of temperature and depending on the available cage-void volume for the MA molecules. The latter, decreases obviously as the halide ionic radius decreases going from I to Br to Cl. In summary, one can make following observations: i) For MAPbI$_3$ the Raman spectra of the cubic and tetragonal phases, where the MA dynamics is unfolded, exhibit very broad peaks and are almost featureless, whereas in the low-temperature orthorhombic phase the Raman peaks are well-defined and much sharper. ii) The other extreme is MAPbCl$_3$ for which the Raman spectra exhibit fairly broad peaks at all temperatures, i.e. for all three phases, although there is a slight but clear temperature dependence of the linewidth. iii) MAPbBr$_3$ is an intermediate case between the other two, being the difference in linewidth of the Raman peaks between the orthorhombic phase and the other two phases much less marked than for MAPbI$_3$.

In the case of MAPbI$_3$, the DSI is the cause of the large inhomogeneous broadening of the inorganic cage phonons in the cubic and tetragonal phases, because of the dynamic disorder induced by the unleashed MA dynamics. In the low-temperature orthorhombic phase, the MA molecules become locked within the cage voids in a state of static order, hence the inhomogeneous broadening vanishes and the linewidths are then determined by the homogeneous part. For MAPbI$_3$ it turns out that the inhomogeneous broadening overwhelms the homogeneous contribution. Since with locked A-site cations there is no DSI, the coupling between both sublattices is solely mediated by H-bonding interaction and the phonon linewidth is just the homogeneous one, determined by the phonon lifetime. The latter is so because, after its transition into the orthorhombic phase, the ferroelastic perovskite material breaks up in macroscopic twin domains \cite{ambro22a}, inside which the MA molecules are perfectly ordered (static order situation). For instance, the use of polarized micro-Raman spectroscopy allowed the direct observation of the parallel alignment of the MA cations in the low-temperature orthorhombic phase of MAPbCl$_3$ \cite{kimxx22a}.  

The disappearance of dynamic disorder and the concomitant vanishing of the inhomogeneous broadening when transforming into the orthorhombic phase is ubiquitous in MHPs, as has been reported from temperature-dependent Raman measurements for MAPbI$_3$ \cite{ivano16a,sharm20a,sharm20b}, MAPbBr$_3$ \cite{guoxx17a,yinxx17a,maczk22a},  MAPbCl$_3$ \cite{naqvi22a}, FAPbBr$_3$ \cite{maczk22a,naqvi23a}, CsPbBr$_3$ \cite{guoxx17a,hoffm23a} and MHyPbCl$_3$ \cite{maczk22b}, where MHy stands for methylhydrazinium. Interestingly, Nakada et al. \cite{nakad19a} has observed a sudden increase in the FWHM also of some molecular modes of the MA cations in MAPbI$_3$ and MAPbBr$_3$ concomitant with the transformation from orthorhombic to tetragonal, i.e. with the unfolding of the MA cation dynamics. In any case, this is a very important result that indicates that, in general, the inhomogeneous broadening of the inorganic cage phonons is greater than the homogeneous linewidth, i.e., $\Delta \omega>\Gamma$ (MAPbCl$_3$ \cite{leguy16a} and FAPbBr$_3$ \cite{maczk22a} are exceptions). This, in turn, provides compelling evidence that, at high temperatures, dynamic disorder, rather than lattice anharmonicity, dominates the coupling between inorganic cage  and A-site cation sublattices. The opposite occurs at low temperatures, when the A-site cations are confined to the cage voids. As shown below, this aspect will be key to the correct interpretation of the "central peak" Raman feature. 

The Raman linewidth can also provide important information about the structural transformations occurring in the MHPs at room temperature but as a function of hydrostatic pressure. Representative examples include MAPbI$_3$ \cite{franc18a} and MAPbBr$_3$ \cite{xuxxx23a}, for which a sudden \textit{sharpening} of the Raman peaks of the inorganic cage modes is clearly observed above 2.7 GPa and ca. 2 GPa, respectively, upon the pressure-induced transformation into orthorhombic phases. According to the previous discussion, such a behavior of the Raman linewidths is an indication that the perovskite transforms from a state of dynamic disorder in the cubic or tetragonal phases, due to an unleashed MA-cation dynamics, to a state of static order with all MA molecules locked inside the cage voids and orderly oriented in the repeated unit cell of the orthorhombic high-pressure phases. The reason why in the high pressure experiments the reduction of the phonon linewidths for MAPbBr$_3$ is not so pronounced as for MAPbI$_3$ is just a consequence of the smaller unit cell volume in the former case. This and the fact that the Raman measurements were carried out at room temperature implies a greater {\it homogeneous broadening} of the Raman peaks due to a stronger coupling with the MA molecules within the narrower voids of the lead bromide cage.

\subsection{\label{Central peak}Origin of the \textit{Central Peak}}

\begin{figure}[H]
  \includegraphics[width=12cm]{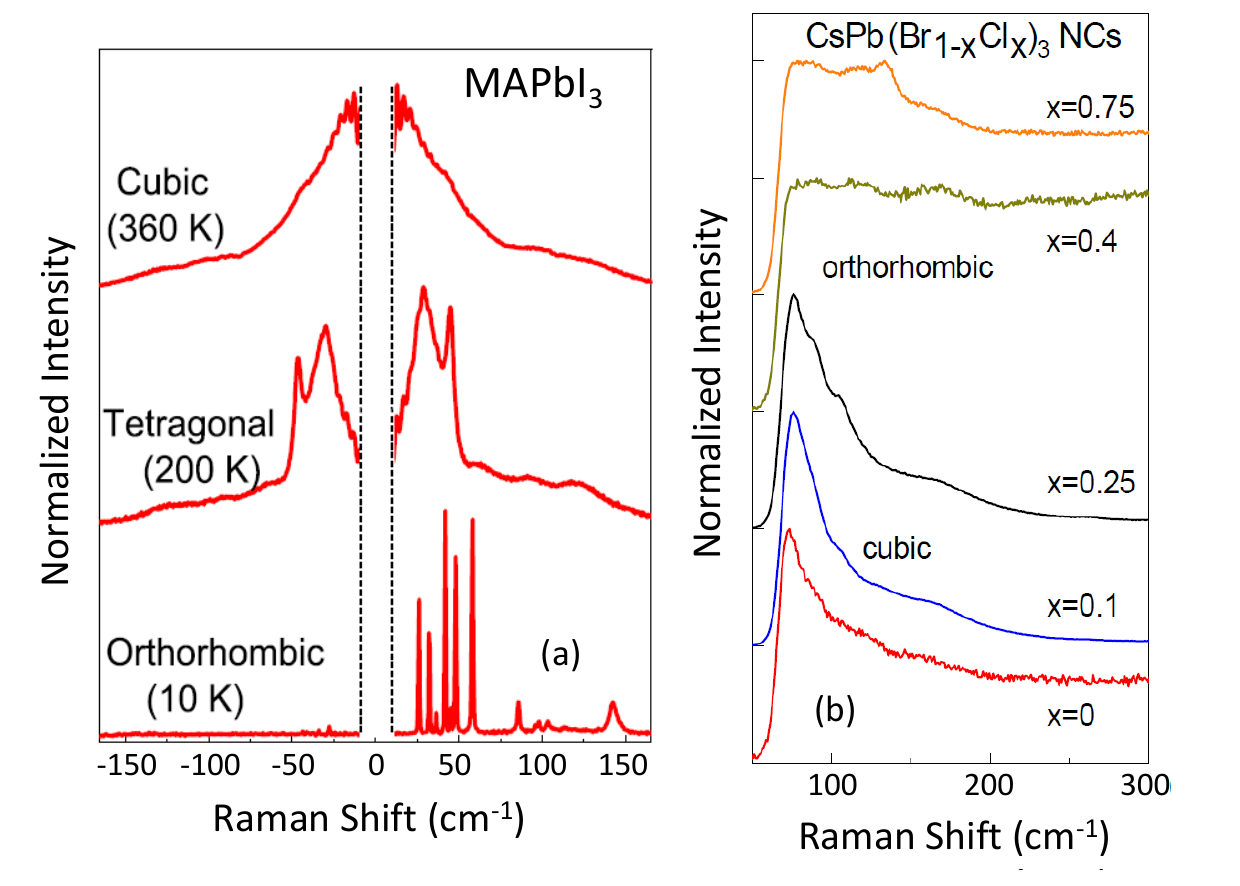}
  \caption{(a) Temperature dependence of low-frequency unpolarized Raman spectra showing the phase transitions in MAPbI$_3$. The Rayleigh scattering region of the spectra is removed from 0 cm$^{-1}$ to the vertical dotted lines (Reproduced with permission from Phys. Rev. Mater. 4, 051601R (2020). Copyright 2020 American Physical Society \cite{sharm20a}). (b) Normalized Raman spectra in the spectral range of the inorganic cage phonons recorded with 785-nm excitation at room temperature for CsPb(Br$_{1-x}$Cl$_x$)$_3$ NCs with chlorine content x=0, 0.10, 0.25, 0.40 and 0.75 (Reproduced with permission from J. Phys. Chem. Lett. 16, 1134 (2025). Copyright 2025 American Chemical Society \cite{fasah25b}). 
  \label{central peak}}
\end{figure}

At this point, it is appropriate to discuss the origin of the aforementioned Raman feature called the central peak \cite{yaffe17a}, taking into account the conclusions drawn so far regarding the relative importance of lattice anharmonicity and dynamic disorder. The central peak corresponds to a background signal, which rises sharply towards a zero Raman shift —hence its name— and is often, but exclusively, observed in the Raman spectra of the cubic and tetragonal phases of MHPs within their stable temperature range. Figure \ref{central peak}a shows three representative Raman spectra of a high-quality MAPbI$_3$ single crystal, recorded in the spectral range of the inorganic cage phonons each for one of the different phases adopted by the material as a function of temperature \cite{sharm20a}. Other examples, where the central peak shows up in Raman, are: again MAPbI$_3$ \cite{nakad19a,sharm20b}, MAPbBr$_3$ \cite{yaffe17a,guoxx17a,nakad19a,maczk22a}, MAPbCl$_3$ \cite{naqvi22a}, FAPbBr$_3$ \cite{maczk22a}, and CsPbBr$_3$ \cite{yaffe17a,guoxx17a}, for instance.  

In part inspired by the extreme similarity with the low-frequency Raman spectrum of liquids \cite{niels79a,perro81a}, the central peak of MHPs was interpreted as arising from light scattering by anharmonic, local polar fluctuations of the perovskite structure \cite{yaffe17a,sharm20b}. In the case of MAPbI$_3$ and with the aid of ab-initio MD calculations, two major sources for the strong anharmonicity leading to the central peak feature were presumably identified, namely, the dynamic disorder caused by orientational unlocking of the MA$^+$ cations and large-amplitude octahedral tilting \cite{sharm20b}. For Cs halide perovskites, in turn, the anharmonicity is supposed to arise from large off-center displacements of the Cs$^+$ cations \cite{yaffe17a}. However, this interpretation presents a conflict. In the previous discussion, it has been shown that when the A-site cation dynamics unfolds (partially or fully), phonon interactions are dominated by DSI and the effects of concomitant (static/dynamic) structural inhomogeneities rather than lattice anharmonics. Furthermore, there is a one-to-one correlation between the freeze-out of the A-site cation dynamics and the \textit{simultaneous} disappearance of the inhomogeneous broadening of the cage-phonon Raman peaks and of the central peak, as observed in any of the examples cited above when the perovskite transforms into an orthorhombic phase at low temperatures. And this is not just an effect of temperature. Figure \ref{central peak}b displays the low-frequency Raman spectra, all taken at room temperature, for a series of CsPb(Br$_{1-x}$Cl$_x$)$_3$ nanocrystals (NCs) with different Cl content \cite{fasah25b}. For Cl concentrations below 40\% the central peak is clearly apparent in the Raman spectra. For these concentrations the NCs are cubic, according to X-ray diffraction, and the Cs dynamics fully unfolded. On the contrary, at high Cl contents the central peak is no longer observable, situation which coincides with the transformation of the NCs into a shrunken orthorhombic phase, for which the Cs cations become locked inside the cage voids \cite{fasah25b}.          
 
Hence, one is led to the conclusion that the appearance of the Raman central peak has to be intimately linked to the presence of structural disorder in the MHP, which is produced when the A-site cation dynamics becomes activated. In view of the fact that the typical timescale of the roto-translational motion of the A-site cations is in the ps range, i.e., much longer than the phonon lifetimes (sub-ps regime) \cite{frost16a}, for the phonons the disorder induced by the A-site cation dynamics is \textit{static} in nature (see discussion of the inhomogeneous broadening above). In this respect, it is very instructive to consider Raman results obtained elsewhere for other semiconductor systems with nanoscale structural disorder and which also show a completely similar phenomenology regarding the observation of a central peak. 

\begin{figure}[H]
  \includegraphics[width=11cm]{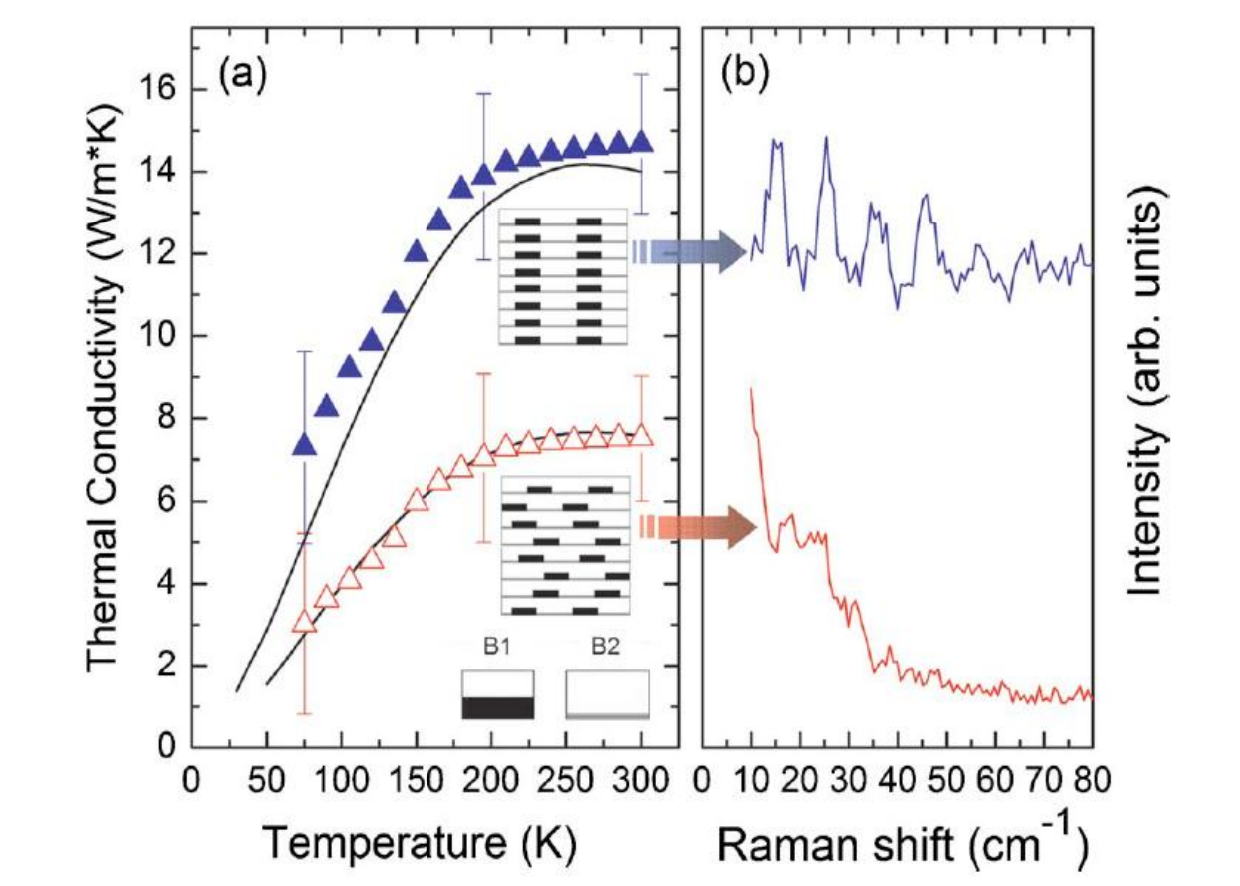}
  \caption{(a) Measured thermal conductivity as a function of temperature for Ge quantum dot superlattice (QDSL) samples with vertically aligned dots (solid blue symbols) and without vertical correlation (open red symbols). Also shown are results of calculations using the extended Fourier heat transport equation (solid line). The inset shows sketches of the QDSL nanostructures with and without vertical correlation. Black squares represent the Ge dots, whereas gray regions correspond to the Ge wetting layer. White regions represent the Si spacer. (b) Measured Raman spectra (Ref. [\cite{lacha07a}]) of the corresponding samples in (a). (Reproduced from J. Alvarez-Quintana et al., Appl. Phys. Lett. 93, 013112 (2008), with the permission of AIP Publishing \cite{alvar08a}).
  \label{Ge QDs}}
\end{figure}

The first example concerns a series of short-period superlattice samples of Ge quantum dots (QDs) embedded in Si, grown by molecular beam epitaxy (MBE) \cite{alvar08a}. These samples were purposely made for studies of the influence of the vertical alignment of the dots in $z$ (growth) direction on the cross-plane thermal conductivity. The vertical correlation between Ge dots can be modified either by proper selection of the spacer-layer thickness or, as done in Ref. [\cite{alvar08a}], by using a seed submonolayer of C, introduced in each period of the multi-stack just before the deposition of the Ge QDs. As shown elsewhere \cite{berna06a}, adding submonolayer amounts of C drastically affects the dot-nucleation mechanism, counteracting the stress memory between layers, thus leading to an almost total loss of vertical correlation. Figure \ref{Ge QDs}a shows an example of two samples grown simultaneously on the same Si substrate, which differ only in the degree of disorder in the vertical alignment of the Ge QDs, exhibiting also a quite different cross-plane thermal conductivity \cite{alvar08a}. Since in these intrinsic (no doping) semiconductor samples the heat is mainly carried by acoustic phonons, the much lower thermal conductivity of the superlattice without vertical alignment of the QDs is indicative of the magnitude of the effect of (static) disorder on the acoustic phonon spectrum.   

Of great relevance to the present discussion is that the effects of disorder are also clearly reflected in Raman spectra in the acoustic phonon spectral range. Figure \ref{Ge QDs}b displays two representative \textit{resonant} Raman spectra of the same QD superlattice samples for which the thermal conductivity was measured, that is one with almost perfect QD alignment in $z$ direction and the other fully disordered \cite{lacha07a}. For the multilayer sample with perfect dot correlation in the growth direction, the interaction of the acoustic phonons with the ensemble of electronic states confined to the dots gives rise to well-defined Raman interferences (coherent scattering). In contrast, the interference contrast almost vanishes when carbon is introduced on the dot-nucleation surfaces in each layer. Instead, a strong and steeply increasing background towards 0 cm$^{-1}$ (the central peak) is observed at small Raman shifts, which is the result of the strongly incoherent acoustic-phonon Raman scattering induced by the lost translational invariance in $z$ direction. These drastic changes in the Raman spectra of QD multilayers with and without C are directly related to the lack or existence of vertical correlation between Ge dots, respectively, as confirmed by calculations within the Raman interference model reported elsewhere \cite{cazay01a,lacha07a}. 
 
The next example is particularly interesting because it also provides a theoretical explanation for both the (spatially) coherent part of the Raman scattering spectra and the incoherent, disorder-induced part, i.e., the background signal eventually related to the central peak. The system under consideration are short-period GaAs/AlAs superlattices (SLs), whose acoustic-phonon Raman spectra show two characteristic features \cite{rufxx94a}: Sharp lines originating in crystal-momentum conserving backscattering by folded SL phonons. A continuous emission background with superimposed peaks and dips observed due to disorder-induced scattering from modes of the whole acoustic-phonon dispersion. Even in high-quality samples grown by MBE the amount of disorder is such that both effects appear at the same time. As for the Ge QDs, the sharp features are typically only observed near (incoming) resonance conditions. Thus, Fig. \ref{GaAs/AlAs SLs}a shows a series of Raman spectra of a 10/10 monolayer GaAs/AlAs SL (closed data points), recorded at 16 K in the spectral range of the acoustic phonons at different detunings $\Delta$ of the incident laser with respect to a direct optical transition of the SL \cite{rufxx94a}. As usual, data points very near the laser line have been omitted, as they are, on purpose, efficiently rejected by the experimental Raman setup. The solid curves correspond to the fitted theoretical acoustic-phonon continuous emission lineshapes, described in the following. 

Neglecting zone-folding effects for acoustic SL phonons, i.e., assuming a linear dispersion for the phonon frequency vs wave vector ($\omega(\mathbf{q})\propto \mathrm{q}$, with $\mathrm{q}^2=\mathrm{q}_z^2 + \mathrm{q}_\|^2$), the Raman intensity for disorder-induced acoustic-phonon
scattering has been demonstrated to be given by \cite{belit94a}:   

\begin{figure}[H]
  \includegraphics[width=13cm]{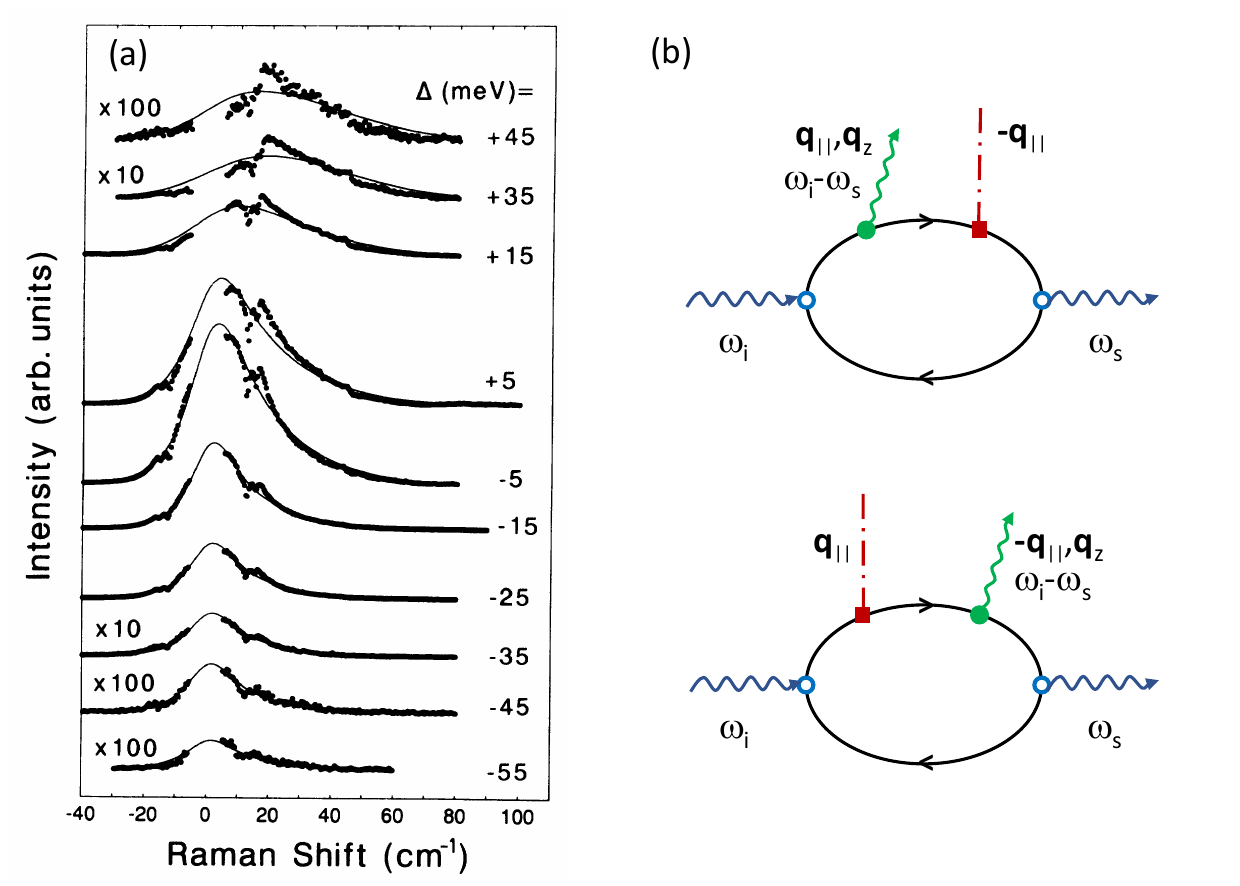}
  \caption{(a) Experimental acoustic-phonon Raman spectra (dots) and theoretical fits (solid lines) of the (10/10) monolayer GaAs/AlAs superlattice for various resonance detunings $\Delta$. Intensity scaling factors are given on the right-hand side of the spectra. A background describing the underlying luminescence has been removed before fitting. Positive Raman shifts indicate Stokes scattering, negative ones anti-Stokes emission (Reproduced with permission from Phys. Rev. B 50, 1792 (1994). Copyright 1994 American Physical Society \cite{rufxx94a}). (b) Representative diagrams of second-order (Stokes) Raman processes with acoustic-phonon (full circle, wavy line) and interface-roughness induced (square, dash-dotted line) scattering of intermediate electronic states.)
  \label{GaAs/AlAs SLs}}
\end{figure}

\begin{equation}
\textsl{I}_R(\omega(\mathbf{q}))\propto\underbrace{\sum_{\mathbf{q}}\left|M(\mathbf{q})\right|^2\delta(\omega_i-\omega_s\mp\omega(\mathbf{q}))\cdot\left(n(\omega(\mathbf{q}))+\frac{1}{2}\pm\frac{1}{2}\right)}_{form factor}\cdot\underbrace{\left|\sum_{m\mathbf{q_\|}} e^{iq_zmd}e^{i\mathbf{q_\|}\bullet\mathbf{r_{xy}}}R(m)\right|^2}_{structure factor},
\label{Form-struc}
\end{equation}
\noindent where the matrix element $M(\mathbf{q})$ contains details of the electron-phonon interaction of acoustic phonons with confined electronic states of individual quantum wells $m$ and the delta function ensures energy conservation, being $\omega_i, \omega_s$ the incident and scattered photon frequency, respectively. Stokes (upper signs) and anti-Stokes (lower signs) processes depend on the Bose-Einstein thermal factor $n(\omega(\mathbf{q}))=1/(e^{\frac{\hslash\omega(\mathbf{q})}{k_BT}}-1)$ with $k_B$ the Boltzmann constant. In Eq. (\ref{denom}), $R(m)$ contains the resonance denominators of the expression for the Raman scattering amplitude which is given by third-order perturbation theory:  

\begin{equation}
\begin{split}
R(m) & = \frac{1}{\left(\Delta(m)+i\gamma\right)\cdot\left(\Delta(m)-\hslash\omega_i+\hslash\omega_s+i\gamma\right)}, \\
\Delta(m) & = \hslash\omega_i - E_m.
\end{split}
\label{denom}
\end{equation}
\noindent Note that $\gamma$ corresponds to the (natural) homogeneous linewidth of the \textit{electronic} interband transitions with energy $E_m$, involved in the Raman scattering process. 

In ideal structures, i.e., with negligible structural disorder, wave vector conservation leads to constructive/destructive interferences of the quantum phases appearing in the sum of the structure factor in Eq. (\ref{Form-struc}). This gives rise to the sharp features (peaks/dips) in the acoustic-phonon Raman spectra of GaAs/AlAs SLs and Ge-QD multistacks, for example. In real structures, disorder causes a distribution of electronic energies and, what is more important, leads to a partial or total breakdown of wave vector conservation. On the one hand, this smears out partially or totally the sharp interference features and, on the other hand, gives rise to a continuous phonon emission/absorption background in the Stokes/anti-Stokes Raman spectra. 

Under resonance conditions, the Raman scattering intensity is dominated by the energy denominator factor $R(m)$ in the structure factor of Eq. (\ref{Form-struc}). Since the disorder-induced variation of the optical transition energies can be modelled by a Gaussian distribution, $R(m)$ thus leads to Gaussian Raman intensity profile as well, centered at zero Raman shift (strict incoming resonance condition). However, the further away from resonance, the contribution from the energy denominator gradually fades and the Raman intensity profile becomes less intense but determined by the form factor in Eq. (\ref{Form-struc}). Assuming wave-vector conservation breakdown, the sum over all wave vectors $\mathbf{q}$ of the form factor has to be replaced by an integral over energy weighted with the phonon density of states (DOS). Thus, the form factor is essentially given by the product of the Bose-Einstein occupation number and the phonon DOS (the  \textit{exact} background lineshape might also depend on $M(\mathbf{q})$\cite{rufxx94a}). For a linear acoustic-phonon dispersion, at small values of $\mathrm{q}$ it holds $n(\omega(\mathbf{q}))\propto \mathrm{q}^{-1}$, while the phonon DOS increases as $\mathrm{q}^2$.  Therefore, their product also exhibits a maximum, but at finite Raman shifts, as illustrated by the measured and fitted Raman spectra of the GaAs/AlAs SL shown in Fig. \ref{GaAs/AlAs SLs}a for different resonance detunings $\Delta$ \cite{rufxx94a}. In this respect, the name of "central peak" for the continuous acoustic-phonon emission background is somewhat misleading. It is worth mentioning that for the computation of the fitting curves in Fig. \ref{GaAs/AlAs SLs}a only fluctuations in layer thickness along the growth direction have been considered and a bulklike linear acoustic-phonon dispersion has been assumed (no folded phonons). This is why they do not exhibit any Raman interference features \cite{rufxx94a}. 

Quantum mechanically, disorder-induced Raman scattering is a second-order (four steps) process. Figure \ref{GaAs/AlAs SLs}b displays two (pseudo) Feynman diagrams for the second order Stokes process of an electron being scattered by a SL acoustic phonon (circle vertex, wavy line) and interacting with the interface-roughness potential (square vertex, dash-dotted line) \cite{belit94a}. The electron-roughness interaction is elastic in nature, which means that the electron, when scattered by the defect or inhomogeneity, conserves energy but acquires the necessary crystal momentum to exactly compensate the momentum change produced by its interaction with the acoustic phonon. This results in the participation in the scattering of acoustic phonons with \textit{all} wave vectors, leading to the continuous acoustic-phonon emission background, which for MHPs is induced by the structural disorder associated with an unfolded A-site cation dynamics. It is worth noting that this theory applies to any type of structural disorder, whether it be fluctuations in the thickness of quantum wells, the alignment of quantum dots, or the disorder caused by the dynamics of the A-site cations.   
 
Finally, a word of caution by the interpretation of the Raman background signal near resonance conditions. In this case, the coexistence of the Raman signal and that stemming from luminescence due to geminate radiative recombination is unavoidable \cite{belit94a,rufxx94a}. In the case of continuous-wave experiments, it is simply not possible to separate the contributions from PL and the continuous phonon emission Raman background, should this be necessary. However, an idea of the relative importance of the PL contribution can be obtained by detuning the resonance. A good example of this can be seen in the Raman spectra taken as a function of temperature for a series of FA$_x$MA$_{1-x}$PbI$_3$ single-crystal samples (see Supplementary Information of Ref. [\cite{franc20a}]). Due to the marked dependence of the band gap on temperature and composition, for excitation with the 785-nm laser line (1.58 eV) the PL signal is so strong that dominates the Raman spectra for FA contents $x\geq0.5$ at all temperatures.

\section{\label{Final}Conclusions}

In summary, it was shown that the interaction between the inorganic cage lattice and the A-site cation network plays a key role in determining phonon interactions in MHPs, depending on whether the inter-sublattice coupling is mediated by hydrogen bonding or by dynamic steric interaction for unleashed A-site cation dynamics. In the former case, H-bonding leads to large lattice anharmonicities which impact the lifetimes of the cage phonons, influencing mainly the homogeneous broadening of the Raman peaks. In contrast, the local structural disorder caused by the roto-translational dynamics of the A-site cations, whether organic or inorganic, leads to strong inhomogeneously broadened Raman peaks and, more importantly, gives rise to the background Raman signal called central peak. By comparison with the observed phenomenology regarding the Raman spectra in the acoustic phonon spectral range for other semiconductor systems exhibiting structural disorder like Ge-QDs multistacks and short-period GaAs/AlAs SLs, the origin of such background was unravelled. Apart from the contribution of geminate PL under near-resonance conditions, the steep background called the central peak is well explained by invoking acoustic-phonon wave vector scrambling in disorder-induced acoustic phonon second-order Raman scattering processes. In this way, a unifying picture of phonon interactions in MHPs is provided, which is instrumental for the correct interpretation of their Raman spectra. This is an issue of great practical importance, due to the relevance of Raman scattering as spectroscopic materials characterization technique.

\section{Acknowledgements}

The author is grateful to Dr. M. I. Alonso and Dr. J. S. Reparaz for a critical reading of the manuscript. The Spanish "Agencia Estatal de Investigación" is gratefully acknowledged for its support through grant CEX2019-000917-S (FUNFUTURE) and CEX2023-001263-S (MATRANS42) in the framework of the Spanish Severo Ochoa Centre of Excellence program and the AEI/FEDER(UE) grants PID2021-128924OB-I00 (ISOSCELLES) and PID2024-163010OB-I00 (PV-MENU). The author also thanks the Catalan agency AGAUR for grant 2021-SGR-00444 and the National Network "Red Perovskitas" (AEI funded).

\end{document}